\documentclass[aps,amsmath,amssymb,twocolumn,longbibliography, superscriptaddress]{revtex4-2}
\usepackage{graphicx}
\usepackage{dcolumn}
\usepackage{bm}
\usepackage{mathrsfs}
\usepackage{subfigure}
\usepackage{natbib}
\usepackage{amssymb}
\usepackage{multirow}
\usepackage{float}
\usepackage{color}
\usepackage{ulem}

\begin{document}

\title{\textcolor{black}{Interference-induced surface superconductivity:\\ enhancement by tuning the Debye energy}}

\author{Yunfei Bai}
\affiliation{Key Laboratory of Optical Field Manipulation of Zhejiang Province, Department of Physics, Zhejiang Sci-Tech University, 310018 Zhejiang, China}

\author{Yajiang Chen}
\email{yjchen@zstu.edu.cn}
\affiliation{Key Laboratory of Optical Field Manipulation of Zhejiang Province, Department of Physics, Zhejiang Sci-Tech University, 310018 Zhejiang, China}

\author{M. D. Croitoru}
\affiliation{HSE University, 101000 Moscow, Russia}

\author{A. A. Shanenko}
\affiliation{HSE University, 101000 Moscow, Russia}

\author{Xiaobing Luo}
\affiliation{Key Laboratory of Optical Field Manipulation of Zhejiang Province, Department of Physics, Zhejiang Sci-Tech University, 310018 Zhejiang, China}

\author{Yunbo Zhang}
\affiliation{Key Laboratory of Optical Field Manipulation of Zhejiang Province, Department of Physics, Zhejiang Sci-Tech University, 310018 Zhejiang, China}

\date{\today}
\begin{abstract}
In the usual perception, surface superconductivity is associated with the surface nucleation of a superconducting condensate above the upper critical field in type-II superconductors or with a rearrangement of phonon properties and the electron-phonon coupling near surfaces/interfaces. Recently, it has been found that there is another example when the surface superconducting temperature is increased up to $20$-$25\%$ as compared to the bulk one due to constructive interference of superconducting pair states. In the present work, we demonstrate that in fact, such an interference-induced enhancement can be much more pronounced, up to nearly $70\%$. Furthermore, here it is shown that such an interference enhancement persists over a wide range of microscopic parameters.
\end{abstract}

\maketitle

\section{Introduction}
\label{int}

There are two well-known examples of the surface superconductivity. The first one concerns the surface nucleation of a pare condensate in type-II superconductors below the third critical field $H_{c3}$, when the applied external magnetic field $H$ is in the interval from $H_{c2}$ to $H_{c3}$, see the pioneering works~\cite{SJames1963, Gennes1964, SJames1965, Gennes1966, SJames1969}. The second variant is related to an enhancement (and also suppression) of superconductivity due to surface modifications of the phonon properties, see e.g. the papers~\cite{Strongin1968, Dickey1968, Naugle1973, Leavens1981}. 

However, there also exists the surface superconductivity enhancement at the zero applied field and without any modifications in the phonon degrees of freedom. For conventional superconductors, the investigations based on both the Ginzburg-Landau (GL) theory~\cite{Fink1965} and the microscopic Bogoliubov-de Gennes (BdG) equations~\cite{Boyd1968, Troy1995, Giamarchi1990} have shown that the order parameter near the surface can be significantly larger than in bulk. This does not necessarily lead to a notable increase of the superconducting transition temperature near the surface $T_{\rm cs}$ as compared to its bulk value $T_{\rm cb}$. The corresponding relative difference between the surface and bulk critical temperatures $(T_{\rm cs}-T_{\rm cb})/T_{\rm cb}$ was reported to be negligible ($\approx 10^{-3}$)~\cite{Giamarchi1990}. However, recently it was found within the BdG equations for the Hubbard attractive model with the nearest-neighbor hopping that the relative difference between $T_{\rm cs}$ and $T_{\rm cb}$ can increase up to $20$-$25\%$, and this increase was attributed to the formation of boundary pair states with elevated critical temperatures~\cite{Samoilenka2020}. Later it was shown~\cite{Croitoru2020, Chen2022} that in fact, the enlargement of the surface critical temperature is caused by the constructive interference of the bulk pair states near the sample surface. Such a constructive interference was found to be most pronounced when the conduction band is symmetric with respect to the Fermi level (the half-filling case).

In the present work we demonstrate that the interference-induced surface superconductivity can result in an even more significant increase of $(T_{\rm cs}-T_{\rm cb})/T_{\rm cb}$, up to $\approx 70\%$. We find that the impact of the interference is notably enhanced by an appropriate tuning of the Debye energy. For illustration, we investigate a one-dimensional (1D) chain of atoms with the $s$-wave pairing of electrons within the tight-binding treatment of the attractive Hubbard model.

The paper is organized as follows. In Sec.~\ref{sec2} we outline the relevant BdG formalism. Section~\ref{sec3} presents our numerical results for $T_{\rm cs}$ and $T_{\rm cb}$ in a wide range of microscopic parameters, such as the Debye energy $\hbar\omega_{\rm D}$, the attractive coupling strength of the Hubbard model $g>0$, and the electron filling number $n_e$. The summary of our results and conclusions are presented in Sec.~\ref{sec4}.  

\section{Theoretical formalism}
\label{sec2}

Let us consider a 1D chain of atoms with the $s$-wave pairing of electrons in the system and adopt the attractive Hubbard model within the tight-binding approximation. The related BdG equations can be written as~\cite{Samoilenka2020, Croitoru2020, Chen2022, Hirsch1992, Tanaka2000}
\begin{align}\label{bdg}
E_\nu u_\nu(i) &= \sum\limits_{i'} H_{ii'} u_\nu(i') + \Delta_i v_\nu(i) \\
E_\nu v_\nu(i) &= \Delta_i^* u_\nu(i) -\sum\limits_{i'} H^*_{ii'} v_\nu(i'),
\end{align}
where $\Delta_i$ is the superconducting order parameter (pair potential) at the lattice site $i$; $H_{ii'}$ is the single-particle Hamiltonian; $E_\nu,\, u_\nu(i)$, and $v_\nu(i)$ are the quasiparticle energy, electron- and hole-like wave functions, respectively. In the absence of external fields, the single-particle Hamiltonian can be written as~\cite{Tanaka2000}
\begin{equation}\label{a}
H_{ii'}=-\sum\limits_\delta t_{\delta}(\delta_{i',i-\delta} + \delta_{i',i+\delta})-\mu\delta_{ii'},
\end{equation}
where $t_\delta$ is the hopping parameter, $\delta$ enumerates the neighboring coupled atomic-like orbitals, $\mu$ is the chemical potential, and $\delta_{i,i'}$ is the Kronecker delta symbol. The Hartree-Fock mean field interaction is ignored here as its main effect is reduced to a shift of the chemical potential, see e.g. Refs.~\onlinecite{Hirsch1992, Chen2009}. 

The BdG equations are solved in the self-consistent manner as $\mu$ and $\Delta_i$ are dependent on the electron- and hole-like wave functions~\cite{Samoilenka2020, Croitoru2020, Hirsch1992, Tanaka2000}. The chemical potential is determined via the equation for the averaged electron filling number (below referred to as the electron density) 
\begin{equation}\label{ne}
n_e = \frac{2}{N}\sum\limits_{\nu,i}\Big\{f_\nu |u_\nu(i)|^2 + (1-f_\nu) |v_\nu(i)|^2\Big\},
\end{equation}
where $f_\nu=f(E_{\nu})$ is the Fermi-Dirac distribution. The site-dependent pair potential $\Delta_i$ is given by  
\begin{equation}\label{op}
\Delta_i = g\sum\limits_{\nu} u_\nu(i)v^*_\nu(i)[1-2f_\nu],
\end{equation}
where the summation is over the BdG pair states $u_{\nu}(i)v^*_{\nu}(i)$ with the quasiparticle energies $0< E_\nu \le \hbar\omega_{\rm D}$~\cite{Gygi1991, Hayashi1998, Matsumoto2001}, where $\omega_D$ is the Debye frequency \textcolor{black}{(for the conventional phonon mediated superconductivity). Here we notice that the superconductive Hubbard model is often used without the energy cutoff as the band width is finite and so, the ultraviolet divergence does not appear. Obviously, this does not distort results when the band width is less than the Debye energy $\hbar\omega_{\rm D}$. However, in the opposite case one should include the ultraviolet cutoff to keep the trace of the phonon characteristic energy and recover the standard BCS results for the parabolic band approximation.}

To solve the BdG equations, we first choose initial values for $\Delta_i$ and $\mu$ and insert them into Eq.~(\ref{bdg}). Second, we derive the quasiparticle energies, electron- and hole-like wave functions by diagonalizing the corresponding BdG matrix. Third, the obtained solutions are plugged in Eqs.~(\ref{ne}) and (\ref{op}) to get new $\Delta_i$ and $\mu$. Then, the procedure is repeated until the convergence is reached. When solving the formalism, we take into account the normalization condition 
\begin{align}
\sum\limits_i \big(|u_\nu(i)|^2 + |v_\nu(i)|^2\big)=1,
\label{nc}
\end{align}
see e.g. Ref.~\onlinecite{Gennes1966}. \textcolor{black}{Notice that $\Delta_i$ can be chosen real in the absence of the magnetic field as the Hamiltonian of the system is time-reversal symmetric.}

In the present work we consider the electron densities $n_e=0.8$-$1.2$. In this case the system is close to the half-filling regime, which steadily guarantees the presence of the surface enhancement of the critical temperature, as shown in the previous work~\cite{Croitoru2020}. The Debye energy and the Hubbard coupling strength are taken as free parameters. To avoid unnecessary complications, we restrict ourselves to the conventional nearest-neighbor approximation, i.e. $\delta=1$ and $t_\delta =t$. \textcolor{black}{Below all the energy related quantities are calculated in units of the hopping parameter $t$, i.e. we set $t=1$.}

Notice that Eq.~(\ref{a}) is written for the case of an infinite chain. To consider the surface enhancement of superconductivity, we investigate a finite 1D chain with infinite potential barriers at the sites $i=0$ and $i=N+1$. The number of atoms contributing to the superconducting condensate is chosen as $N=301$, which is sufficiently large to avoid any quantum-size effects. For such a finite 1D chain one should keep in mind that the first term in the parenthesis of the right-hand side of Eq.~(\ref{a}) is multiplied by $1-\delta_{i,0}$ whereas the second term is multiplied by $1-\delta_{i,N+1}$. In addition, we have the boundary conditions 
\begin{align}
u_{\nu}(0)=u_{\nu}(N+1)=0,\;v_{\nu}(0)=v_{\nu}(N+1)=0.
\label{bc}
\end{align}
This, taken together with Eq.~(\ref{op}), results in $\Delta_0 = \Delta_{N+1}=0$.

\begin{figure}[t]
\resizebox{1\columnwidth}{!}{\rotatebox{0}{\includegraphics{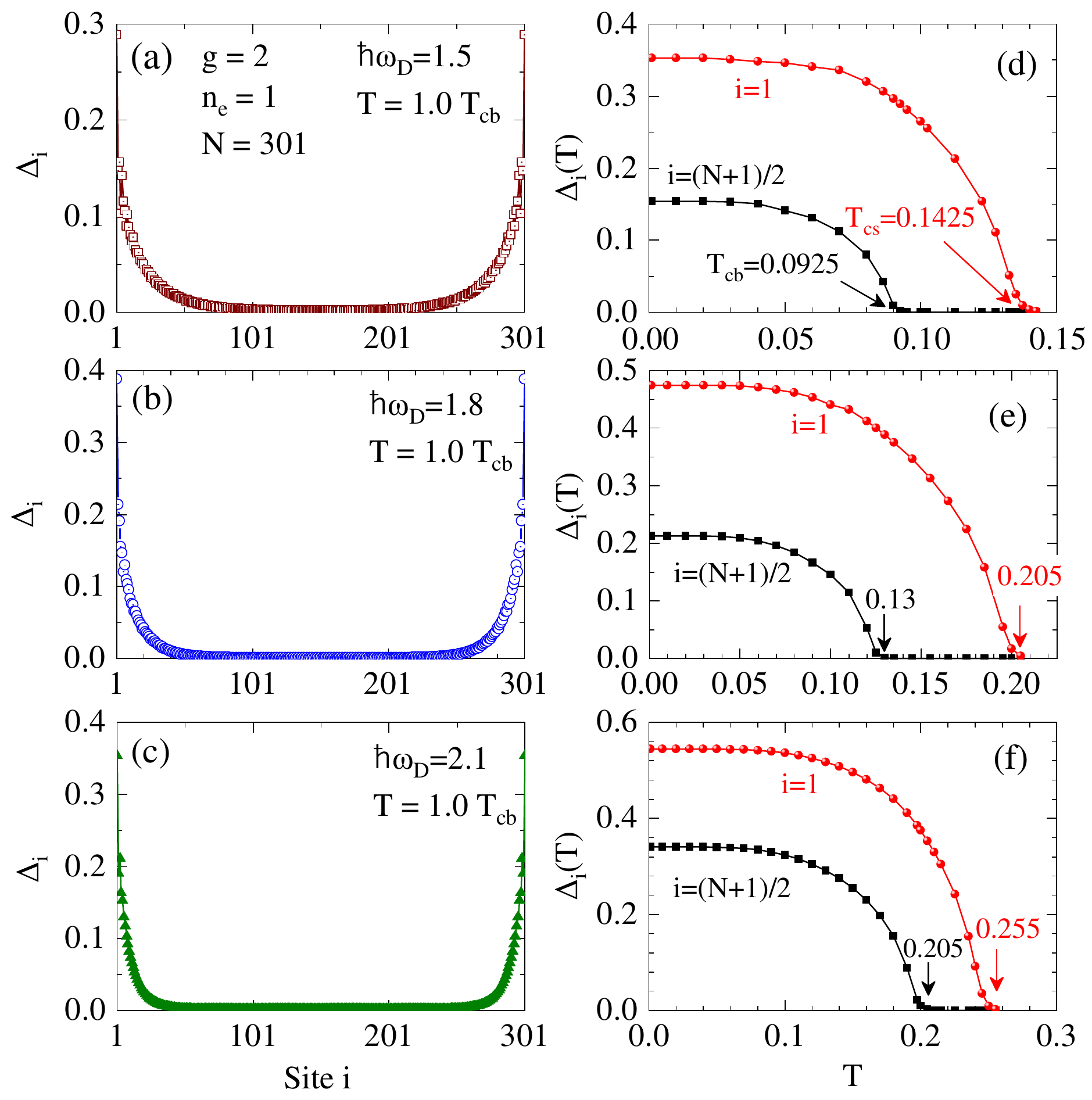}}}
\caption{(a,b,c) The pair potential $\Delta_i$ versus the site number $i$, calculated at the bulk critical temperature. (d,e,f) The pair potentials at the edge (surface) $\Delta_1$ and in the center of the chain (bulk) $\Delta_{(N+1)/2}$ versus the temperature. The calculations are done at $\hbar\omega_{\rm D} = 1.5$~(a),(d), $1.8$~(b),(e), and $2.1$~(c),(f) for $g=2$ and $n_e=1$, other parameters are discussed in the text.}
\label{fig1}
\end{figure}

\section{Results and discussions}
\label{sec3}

\subsection{Surface superconductivity}

Figures~\ref{fig1}(a)-(c) show the order parameter $\Delta_i$ calculated at the bulk critical temperature $T=T_{\rm cb}$ for $n_e=1, g=2$, and the three values of the Debye energy $\hbar\omega_{\rm D}=1.5$~(a), $1.8$~(b) and $2.1$~(c). [We recall that all the energy related quantities are given in units of the hopping parameter $t$.] In these plots, $\Delta_i$ vanishes in the center of the chain (bulk) while it is finite near the edges (surface). As is seen, the system exhibits the surface enhancement of superconductivity. The values of $\Delta_1=\Delta_N$ are sensitive to the Debye energy. For $\hbar\omega_{\rm D}=1.5$ we have $\Delta_1 = 0.29$ whereas for $\hbar\omega_{\rm D}=1.8$ and $\hbar\omega_{\rm D}= 2.1$ we obtain $\Delta_1 = 0.39$ and $\Delta_1=0.35$.  

\begin{figure}[tb]
\resizebox{0.7\columnwidth}{!}{\rotatebox{0}{\includegraphics{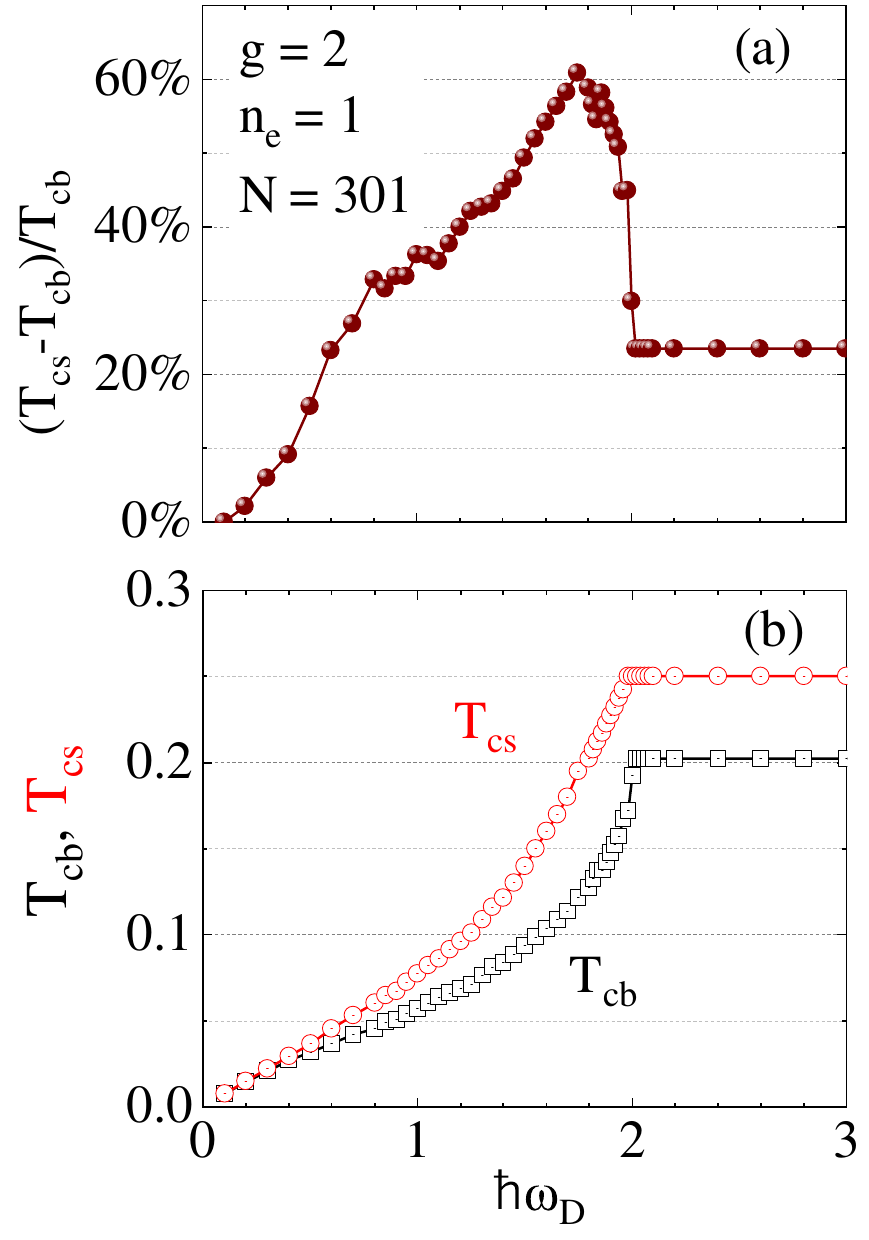}}}
\caption{(a) The difference of $T_{\rm cs}$ and $T_{\rm cb}$ in units of $T_{\rm cb}$ as a function of $\hbar\omega_{\rm D}$. (b) $T_{\rm cs}$ and $T_{\rm cb}$ versus $\hbar\omega_{\rm D}$. The microscopic parameters are the same as in Fig.~\ref{fig1}.} \label{fig2}
\end{figure}

For further details, Figs.~\ref{fig1}(d)-(f) demonstrate $\Delta_1$ and $\Delta_{(N+1)/2}$~(bulk) as functions of the temperature $T$ for $\hbar\omega_{\rm D}=1.5$, $1.8$, and $2.1$, respectively. The electron density and the coupling strength are the same as in Figs.~\ref{fig1}(a)-(c). One sees that $\Delta_1$ and $\Delta_{(N+1)/2}$ approach zero at different temperatures, which is in agreement with the data shown in Fig.~\ref{fig1}(a)-(c). Thus, in addition to the bulk critical temperature $T_{\rm cb}$, associated with the temperature dependence of $\Delta_{(N+1)/2}$, there exists the surface critical temperature $T_{\rm cs}$, associated with the temperature behavior of the edge order parameter $\Delta_1$. 

The both critical temperatures $T_{\rm cb}$ and $T_{\rm cs}$ increase with $\hbar\omega_{\rm D}$: for $\hbar\omega_{\rm D}=1.5$, $1.8$ and $2.1$, we have $T_{\rm cb} = 0.0925$, $0.13$ and $0.205$ and $T_{\rm cs}=0.1425$, $0.205$ and $0.255$, respectively. However, $T_{\rm cb}$ and $T_{\rm cs}$ are not simply proportional to $\hbar\omega_{\rm D}$ as in the conventional BCS model. This is clearly seen from Fig.~\ref{fig2}, where $(T_{\rm cs}-T_{\rm cb})/T_{\rm cb}$ and $T_{\rm cs},T_{\rm cb}$ are shown versus the Debye energy in panels (a) and (b). The calculations are done at $g=2$ for the half-filling case, similarly to Fig.~\ref{fig1}. If $T_{\rm cs}$ and $T_{\rm cb}$ were proportional to $\hbar\omega_{\rm D}$, the relative difference between $T_{\rm cs}$ and $T_{\rm cb}$ in Fig.~\ref{fig2}(a) would be constant for any value of the Debye energy. However, $(T_{\rm cs}-T_{\rm cb})/T_{\rm cb}$ exhibits a complex nonmonotonic dependence on the Debye energy when $\hbar\omega_{\rm D} < 2$ and becomes constant only when $\hbar\omega_{\rm D}$ exceeds $2$. From Fig.~\ref{fig2}(b) one can see that $T_{\rm cb}$ and $T_{\rm cs}$ are almost linear in $\hbar\omega_{\rm D}$ only for $\hbar\omega_{\rm D} \lesssim 0.4$. In the region $0.4 < \hbar\omega_{\rm D} \le 1.75$ the trend becomes different: both $T_{\rm cs}$ and $T_{\rm cb}$ start to rise with $\hbar\omega_{\rm D}$ much faster. Furthermore, $T_{\rm cs}$ increases with $\hbar\omega_{\rm D}$ faster than $T_{\rm cb}$, which leads to the notable increase of the relative difference between $T_{\rm cs}$ and $T_{\rm cb}$, see Fig.~\ref{fig2}(a). Then, near $\hbar\omega_{\rm D}=2$ both critical temperatures approach their maximal values $T_{\rm cs,max}=0.25$ and $T_{\rm cb,max}=0.202$. As a result, the relative difference of the surface and bulk critical temperatures first reaches its maximum of about $61\%$ at $\hbar\omega_{\rm D}=1.75$ and then, drops to the value $(T_{\rm cs,max}-T_{\rm cb,max})/T_{\rm cb,max}=23.4\%$ at $\hbar\omega_{\rm D}=2$. For larger values of the Debye energy the relative difference of $T_{\rm cs}$ and $T_{\rm cb}$ remains $23.4\%$. 

To get an insight into the results in Fig.~\ref{fig2}, let us consider the system at temperatures $T \sim T_{\rm cs}$. In this case the order parameter is sufficiently small and the quasiparticle energy approaches the absolute value of the single-particle energy $\xi_k$~(absorbing the chemical potential). For the single-particle Hamiltonian given by Eq.~(\ref{a}) with the nearest-neighbor hopping, one obtains~\cite{Tanaka2000} 
\begin{equation}
\xi_k=-2{\rm cos}(ka) -\mu,
\label{xi}
\end{equation}
with $a$ the distance between the neighboring sites of the 1D chain and $k$ the crystal momentum. For the half-filling case $\mu=0$ and the modulus of the single-particle energy spans the interval from $0$ to $2$ and so does the quasiparticle energy at $T \sim T_{\rm cs}$. According to the selection rule of Eq.~(\ref{op}), only the BdG pair states corresponding to the quasiparticle energies smaller than $\hbar\omega_{\rm D}$ should be taken into consideration. Then, for relatively small Debye energies, the order parameter includes the BdG pairs states with $0 < E_{\nu} < \hbar\omega_{\rm D} < 2$. In this case the order parameter and the both critical temperatures should increase with the Debye energy because a larger number of the states is incorporated. This increase becomes more pronounced when the Debye energy approaches $2$ and nearly degenerate BdG pair states associated with the edges of the Brillouin zone come into play. However, when the Debye energy exceeds the band width, i.e. $\hbar\omega_{\rm D}>2$, a further increase of $\hbar \omega_{\rm D}$ does not produce any effect on the superconducting properties since all possible pair states are already taken into account. This is why $T_{\rm cs}$ and $T_{\rm cb}$ in Fig.~\ref{fig2} do not change with the Debye energy for $\hbar\omega_{\rm D} > 2$. We stress that this conclusion is only related to the half-filling case with $\mu=0$. For $n_e < 1$ or $n_e >1$ the chemical potential deviates from $0$, and the maximal energy of the contributing quasiparticles becomes larger than $2$, see our results discussed below.  

\subsection{Interference of the BdG pair states}

It is explained in the previous subsection why $T_{\rm cs}$ and $T_{\rm cb}$ increase with the Debye energy for $\hbar\omega_{\rm D} \le 2$ while remaining the same for $\hbar\omega_{\rm D} \gtrsim 2$. However, those arguments cannot explain why we have the surface critical temperature $T_{\rm cs} > T_{\rm cb}$. From the earlier work~\cite{Croitoru2020} we know that the effect of the surface enhancement of superconductivity comes from the constructive interference of the BdG pair states near the surface (edge) of the system. Exactly this constructive interference results in the appearance of the surface critical temperature rather than any superconducting pair mode localized near the edges of the chain. This feature has been revealed in Ref.~\onlinecite{Croitoru2020} for $\hbar\omega_{\rm D} \gg 2$, when all the solutions of the BdG equations contribute to the pair potential~(\ref{op}) and the analysis of their contributions is not complicated by the application of the selection rule for the quasiparticle energies. However, it follows from our present results that the surface enhancement is much more pronounced for the Debye energies in the interval from $1.5$ to $2.0$ which was not investigated in Ref.~\onlinecite{Croitoru2020}. To fill this gap, below we analyze the contributions of the BdG pair states $u_\nu(i)v^*_\nu(i)$ to the order parameter near the edges of the 1D chain and in its center for $\hbar\omega_D \lesssim 2$.

\begin{figure}[tb]
\resizebox{0.65\columnwidth}{!}{\rotatebox{0}{\includegraphics{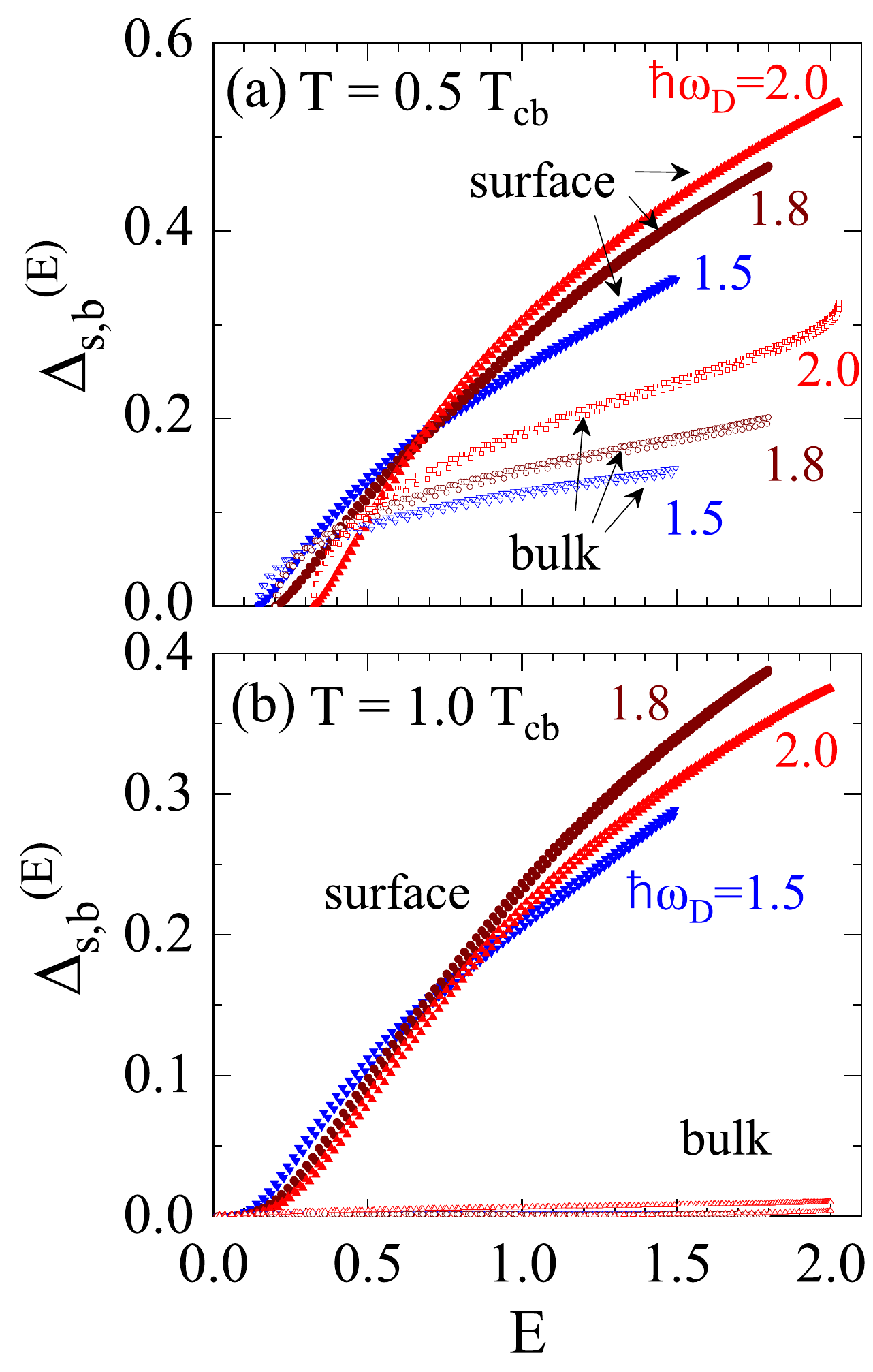}}}
\caption{The cumulative pair potentials $\Delta^{(E)}_s\equiv \Delta^{(E)}_1$ and $\Delta^{(E)}_b\equiv \Delta^{(E)}_{(N+1)/2}$ calculated for $\hbar\omega_{\rm D} = 1.5$, $1.8$ and $2$. Panel (a) corresponds to $T=0.5\,T_{\rm cb}$, and panel (b) is for $T=1.0\,T_{\rm cb}$. Other parameters are the same as in Figs.~\ref{fig1} and \ref{fig2}.} \label{fig3}
\end{figure}

In particular, we follow the paper~\cite{Croitoru2020} and investigate the quantity [for $i=1$ and $i=(N+1)/2$]
\begin{equation}
\Delta_i^{(E)} = g\sum_{0<E_\nu\le E} u_\nu(i)v^*_\nu(i)[1-2f(E_\nu)],
\label{cumop}
\end{equation}
below referred to as the cumulative pair potential (order parameter). Figures~\ref{fig3}(a) and (b) demonstrate the surface cumulative pair potential $\Delta_{\rm s}^{(E)} \equiv \Delta_1^{(E)}$ and the bulk cumulative order parameter $\Delta_{\rm b}^{(E)} \equiv \Delta_{(N+1)/2}^{(E)}$ at $T=0.5T_{\rm cb}$ and $T_{\rm cb}$. As $\Delta^{(E=\hbar\omega_{\rm D})}_i =\Delta_i$, the data in Figs.~\ref{fig3}(a) and (b) are shown for $E \le \hbar\omega_{\rm D}$; the minimal $E$ for non-zero $\Delta_{\rm s}^{(E)}$ and $\Delta_{\rm b}^{(E)}$ corresponds to the lowest quasiparticle energy (the energy gap). The upper three curves in both panels of Fig.~\ref{fig3} are the results for $\Delta_{\rm s}^{(E)}$ at $\hbar\omega_{\rm D}=1.5$, $1.8$, and $2$ while the lower three curves represent $\Delta_{\rm b}^{(E)}$ calculated for the same values of the Debye energy. In the calculations we use $g=2$ and $n_e=1$, similarly to Figs.~\ref{fig1} and \ref{fig2}. 

As is seen from Fig.~\ref{fig1}(a), $\Delta_{\rm s}^{(E)}$ and $\Delta_{\rm b}^{(E)}$ are close to one another for small $E$~($\Delta_{\rm b}^{(E)}$ is only slightly larger). With increasing $E$, the trend changes so that the contribution of the pair states to the surface cumulative order parameter becomes larger than their contribution to $\Delta_{\rm b}^{(E)}$. For example, for $\hbar\omega_{\rm D}=1.8$ this occurs at $E > 0.4$ while for $\hbar\omega_{\rm D}=2.0$ the trend changes above $E =0.5$. One can see that there are no pair states that contribute to $\Delta_{\rm s}^{(E)}$ but do not make any contribution to $\Delta_{\rm b}^{(E)}$ at $T=0.5T_{\rm cb}$. This analysis clearly demonstrates that the surface amplification of the superconducting critical temperature is a consequence of near-surface constructive interference between the pair states spanning the entire system volume, and not a correlation between electrons in localized surface states.

We also cannot find any particular state which makes a major contribution to $\Delta_{\rm s}^{(E)}$ at $T=T_{\rm cb}$. As is seen from Fig.~\ref{fig3}(b), all solutions of the BdG equations with $E_{\nu} \le E$ contribute to $\Delta_{\rm s}^{(E)}$ and so, $\Delta_1$ is controlled by all pair states with $E_{\nu} \le \hbar\omega_D$. Thus, we conclude that the constructive interference of the BdG pair modes is responsible for a nonzero superconducting condensate near the chain edges at the bulk critical temperature (and above $T_{\rm cb}$). This finding is similar to the earlier results~\cite{Croitoru2020} obtained for the Debye energies significantly larger than the band width $\hbar\omega_{\rm D} \gg 2$. 

\textcolor{black}{Based on the interference scenario of the surface enhancement of superconductivity, the appearance of the maximum of $(T_{\rm cs}-T_{\rm cb})/T_{\rm cb}$ as a function of the Debye energy can be explained as follows. At $\hbar\omega_D=0$ we have $T_{\rm cs}=T_{\rm cb}=0$ and so, the relative difference of the surface and bulk critical temperatures is equal to zero. As the Debye frequency increases, more and more pair states appear that contribute to the superconducting condensate. Obviously, the presence of a significant number of participating pair states is necessary for a pronounced constructive interference of such states. This is why the interference effect gets stronger as $\hbar\omega_D$ increases. However, when the number of pair states contributing to the gap function becomes very large, the interference may suffer from an almost random summation of a large number of different terms (similarly to the random phase approximation). This suggests that the surface effect should be maximum at a certain value of $\hbar\omega_{\rm D}$, which is in agreement with our results for $(T_{\rm cs}-T_{\rm cb})/T_{\rm cb}$ shown in Fig.\ref{fig2}(b).}

\begin{figure}[htp]
\resizebox{1\columnwidth}{!}{\rotatebox{0}{\includegraphics{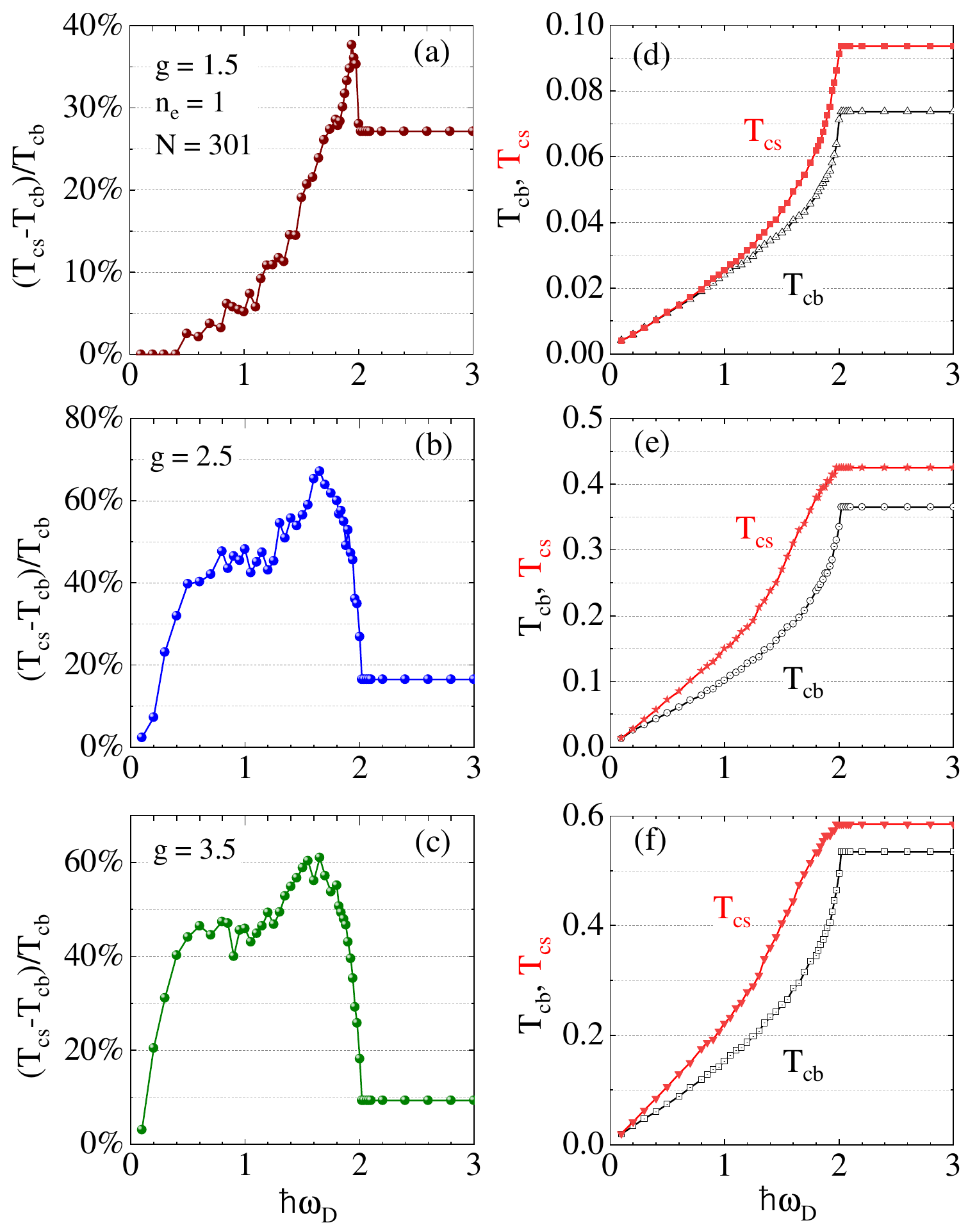}}}
\caption{$(T_{\rm cs}-T_{\rm cb})/T_{\rm cb}$, $T_{\rm cs}$, and $T_{\rm cb}$ as functions of $\hbar\omega_{\rm D}$ at $g=1.5$ (a,d), $2.5$ (b,e) and $3.5$ (c,f) for the half-filling case.} \label{fig4}
\end{figure}

\subsection{Relative difference of $T_{\rm cs}$ and $T_{\rm cb}$ as a function of microscopic parameters}

Here we investigate how the surface enhancement of superconductivity is sensitive to the coupling $g$ and electron density $n_e$. In Fig.~\ref{fig4} one can find $(T_{\rm cs}-T_{\rm cb})/T_{\rm cb}$~(a,b,c) and $T_{\rm cs},T_{\rm cb}$~(d,e,f) as functions of the Debye energy calculated for the half-filling case and the couplings $g=1.5$~(a,d), $g=2.5$~(b,e), and $g=3.5$~(c,f).

From Fig.~\ref{fig4}(a), we find that for $g=1.5$ the maximal relative difference between $T_{\rm cs}$ and $T_{\rm cb}$ is about $38\%$, which is by a factor of $1.4$ larger than its value $27\%$ for $\hbar\omega_{\rm D} >2$. For small values of the Debye energy from $0$ to $\approx 0.5$ the quantity $(T_{\rm cs}-T_{\rm cb})/T_{\rm cb}$ is zero or nearly zero since $T_{\rm cs}$ and $T_{\rm cb}$ approach each other for $\hbar\omega_{\rm D} \to 0$, see Fig.~\ref{fig4}(b). The relative difference between the surface and bulk critical temperatures starts to sharply increase with $\hbar\omega_{\rm D}$ only when the Debye energy exceeds $1.5$ and then, the maximum of $(T_{\rm cs}-T_{\rm cb})/T_{\rm cb}$ is reached at $\hbar\omega_{\rm D}=1.93$.   

The results change significantly for larger couplings. In particular, one can see from Fig.~\ref{fig4}(b) that for $g=2.5$, the relative difference of the surface and bulk critical temperatures can increase up to $67\%$, which is much larger than the maximal value of this quantity at $g=1.5$~($38\%$). Furthermore, $67\%$ is about $4$ times larger than the value of $(T_{\rm cs}-T_{\rm cb})/T_{\rm cb}$ for $\hbar\omega_{\rm D} >2$ at the same coupling ($16\%$). In addition, here the relative difference of $T_{\rm cs}$ and $T_{\rm cb}$ begins to rapidly increase with the Debye energy when $\hbar\omega_{\rm D}$ crosses $0.1$, which is much smaller than $\hbar\omega_{\rm D}=1.5$, the onset of such an increase for $g=1.5$. One can see that there are two intervals where $(T_{\rm cs}-T_{\rm cb})/T_{\rm cb}$ exhibits a significant growth: from $0.1$ to $0.3$ and from $1.1$ to $1.6$. For $\hbar\omega_{\rm D}= 0.3$-$1.1$ we have a saturation of this quantity near $43\%$ whereas its maximal value is reached at $\hbar\omega_{\rm D}=1.6$.  

While the data for $(T_{\rm cs}-T_{\rm cb})/T_{\rm cb}$ at $g=2.5$ are significantly different from those of $g=1.5$, the relative difference between $T_{\rm cs}$ and $T_{\rm cb}$ calculated at $g=3.5$ and shown in Fig.~\ref{fig4}(c) is close to the result for this quantity given in Fig.~\ref{fig4}(b). The only minor difference is that the values of $(T_{\rm cs}-T_{\rm cb})/T_{\rm cb}$ in panel (c) are by about $6\%$ smaller than those in panel (b) for large $\hbar\omega_{\rm D}$. However, $T_{\rm cs}$ and $T_{\rm cb}$ shown in Fig.~\ref{fig4}(f) are more significantly different from the critical temperatures given in Fig.~\ref{fig4}(e). Although the qualitative picture of the Debye-energy dependence of $T_{\rm cs}$ and $T_{\rm cb}$ is the same in both panels, $T_{\rm cs,max}$ for $g=3.5$ is larger by about $30\%$ than $T_{\rm cs,max}$ for $g=2.5$. A similar result is obtained for $T_{\rm cb}$. 

Thus, we find that the maximal value of the relative difference between $T_{\rm cs}$ and $T_{\rm cb}$ at $n_e=1$ increases with $g$ at small couplings, then approaches almost $70\%$ at $g \approx 2.5$, and slowly decreases with a further increase in $g$. Furthermore, one sees in Figs.~\ref{fig3}(a)-(c) that $(T_{\rm cs}-T_{\rm cb})/T_{\rm cb}$ is above $40\%$ in a wide range of the microscopic parameters $0.4 <\hbar\omega_{\rm D}<1.9$ and $2.0 \lesssim g \lesssim 3.5$.

\begin{figure}[tb]
\resizebox{1\columnwidth}{!}{\rotatebox{0}{\includegraphics{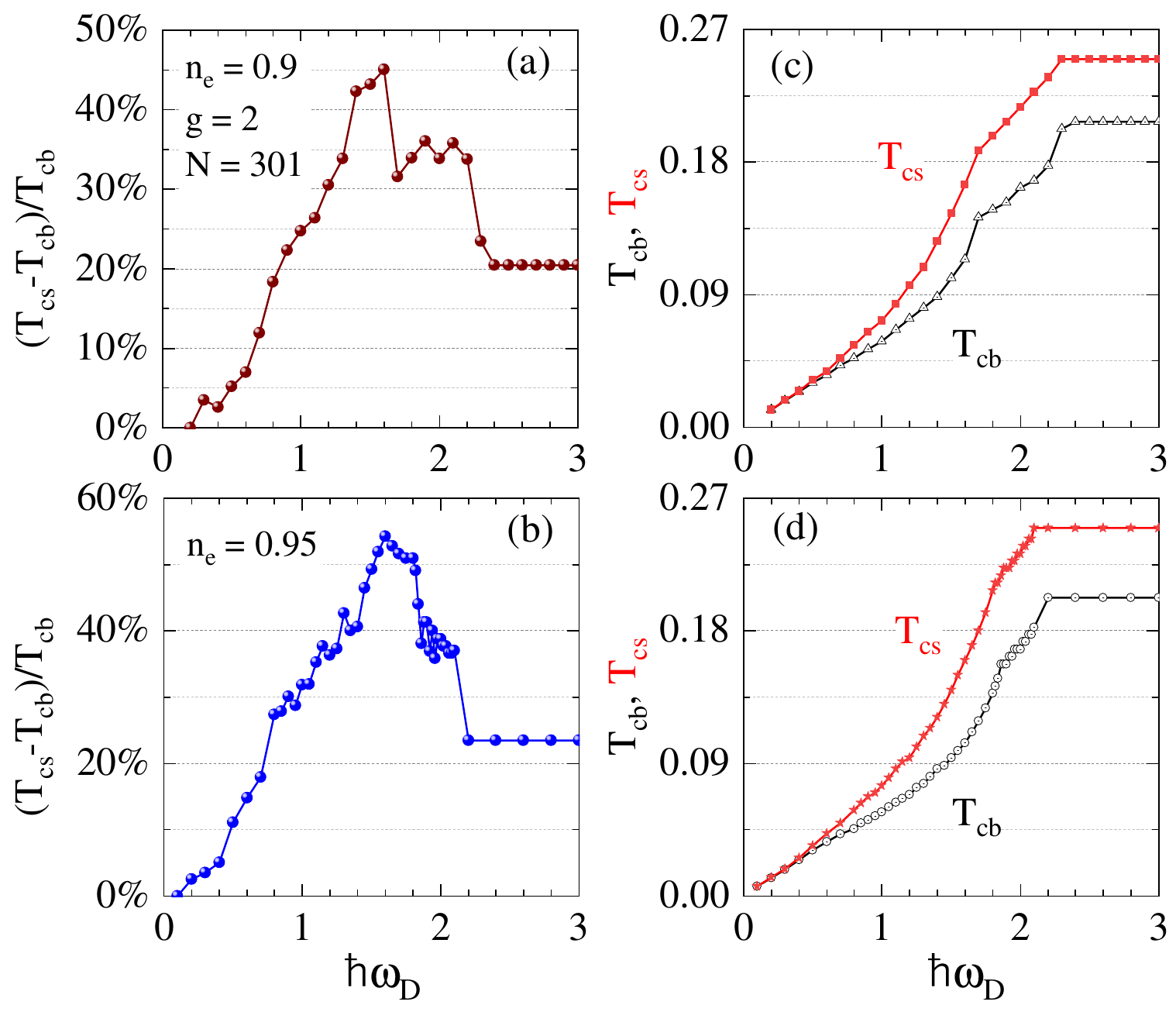}}}
\caption{Beyond the half-filling: $(T_{\rm cs}-T_{\rm cb})/T_{\rm cb}$, $T_{\rm cs}$, and $T_{\rm cb}$ as functions of $\hbar\omega_{\rm D}$ at $n_e=0.9$~(a,c) and $n_e=0.95$~(b,d). The coupling is chosen as $g=2$.} \label{fig5}
\end{figure}

Finally, we go beyond the half-filling regime and investigate how $T_{\rm cs}$, $T_{\rm cb}$ and their relative difference depend on $\hbar\omega_{\rm D}$ at the densities $n_e=0.9$ and $0.95$. Due to the symmetry of the Hubbard model, the results for $n_e=0.9$ and $0.95$ are the same as for $n_e=1.1$ and $1.05$, respectively.

Figures~\ref{fig5}(a,c) and (b,d) demonstrate $(T_{\rm cs}-T_{\rm cb})/T_{\rm cb}$, $T_{\rm cs}$, and $T_{\rm cb}$ as functions of $\hbar\omega_{\rm D}$ calculated for $g=2$ at $n_e=0.9$~(a,c) and $n_e=0.95$~(b,d). One sees from Fig.~\ref{fig5} that for $n_e=0.9$ and $n_e=0.95$ the maximal relative difference between $T_{\rm cs}$ and $T_{\rm cb}$ is about $45.0\%$ and $54\%$, respectively, which should be compared with the maximal relative difference ($60\%$) in Fig.~\ref{fig2}(a) for the same coupling $g=2$. The locus of the maximum of $(T_{\rm cs}-T_{\rm cb})/T_{\rm cb}$ is at $\hbar\omega_{\rm D}=1.60$ for the both densities. When the density $n_e$ is shifted further to $0.85$ and $0.8$, the surface enhancement of superconductivity continues to slightly weaken so that the maximal relative difference between $T_{\rm cs}$ and $T_{\rm cb}$ approaches $38.0\%$ and $29.0\%$, respectively. These results are in agreement with the conclusions of Ref.~\onlinecite{Croitoru2020} that the interference surface effect is most pronounced in the half-filling regime. However, the decrease of $(T_{\rm cs}-T_{\rm cb})/T_{\rm cb}$ calculated at $n_e < 1 (>1)$ with respect to its value at $n_e=1$ is moderate. For example, in the density interval from $0.8$ to $1.2$, we obtain for $g=2$ that the maximal relative enhancement of the surface critical temperature is above $29\%$. Notice that this is still larger than the surface enhancement obtained for the half-filling regime at $\hbar\omega_{\rm D} > 2$ in Refs.~\onlinecite{Samoilenka2020} and \onlinecite{Croitoru2020}.

It is seen from Figs.~\ref{fig5}(c) and (d) that $T_{\rm cs}$ and $T_{\rm cb}$ for $n_e=0.9$ are nearly the same as the critical temperatures calculated for $n_e=0.95$. Qualitatively, their Debye-energy dependence is similar to that demonstrated in Fig.~\ref{fig4} for the half-filling case. However, there is a new feature to discuss: $T_{\rm cb}$ and $T_{\rm cs}$ exhibit the presence of cusps situated at $\hbar\omega_{\rm D}=1.7$ for $n_e=0.9$ and at $\hbar\omega_{\rm D}=1.8$ for $n_e=0.95$~(for both $T_{\rm cb}$ and $T_{\rm cs}$). The reason for the formation of these cusps is the following. At sufficiently large temperatures we can assume that the quasiparticle energy approaches the modulus of the single-particle energy given by Eq.~(\ref{xi}). As $n_e < 1$, the chemical potential $\mu$ is not any more in the center of the band but shifts down, i.e. $\mu<0$. We can distinguish the two branches with $\xi_k >0$ and $\xi_k \le 0$. When the Debye energy is smaller than $|\xi_{k=0}|=2+\mu$ and increases, new contributing BdG states are supplied by the both branches. However, when $\hbar\omega_{\rm D}$ exceeds $|\xi_{k=0}|$, the increase of $T_{\rm cs}$ and $T_{\rm sc}$ occurs only due to the BdG states with $\xi_k >0$ and as a result, the cusps in the Debye-energy dependence of $T_{\rm cs}$ and $T_{\rm cb}$ appear. For $n_e=0.9$ they appear at $\hbar \omega_{\rm D}=1.7$ since $|\xi_{k=0}|=1.7$ and $\mu=-0.3$. In turn, for $n_e=0.95$ one gets $\mu=-0.2$, and the cusps are situated at $\hbar \omega_{\rm D}=1.8$. For the half-filling case their locus approaches $\hbar\omega_{\rm D}=2$. 

Obviously, the maximal values of $T_{\rm cb}$ and $T_{\rm cs}$ are reached when the Debye energy exceeds the value $|\xi_{k=\pi/a}|=2-\mu$ for $\mu\leq 0$. For the half-filling case $\mu=0$ and $|\xi_{k=\pi/a}|=2$. Then, the both critical temperatures approach their maxima at $\hbar\omega_{\rm D}=2$, see Figs.~\ref{fig4}(d,e) and (f). For $n_e=0.95$ we have $\mu=-0.2$ and $|\xi_{k=\pi/a}|=2.2$ while for $n_e=0.9$ one obtains $\mu=-0.3$ and $|\xi_{k=\pi/a}|=2.3$. Therefore, the relative difference of the surface and bulk critical temperatures does not change when $\hbar\omega_{\rm D}$ becomes larger than $2.2$ and $2.3$, respectively.

\section{Conclusions and discussions}
\label{sec4}

In summary, we find that tuning the Debye frequency has a significant impact on the surface enhancement of superconductivity in the attractive Hubbard model with the nearest-neighbor hopping (for the phonon-mediated superconductivity). In particular, our study reveals that $(T_{\rm cs}-T_{\rm cb})/T_{\rm cb}$ can increase up to nearly $60$-$70\%$ for the Debye energies in the interval $1.6$-$1.8$~(in units of the hopping parameter). This is significantly larger than $20$-$25\%$ reported previously for the same model with $\hbar\omega_{\rm D}\geq 2$~\cite{Samoilenka2020, Croitoru2020}. 

We demonstrate that a pronounced surface enhancement of superconductivity persists over a wide range of the microscopic parameters. Indeed, the effect is not very sensitive to a particular value of the coupling constant in the interval $2$-$3.5$ where the maximum of $(T_{\rm cs}-T_{\rm cb})/T_{\rm cb}$ is about $60$-$70\%$ for the half-filling case. When the system deviates from the half-filling regime, the maximum of $(T_{\rm cs}-T_{\rm cb})/T_{\rm cb}$ decreases in agreement with findings in Ref.~\onlinecite{Croitoru2020}. However, it remains significant. For example, at $g=2$ the maximal value of the relative difference between $T_{\rm cs}$ and $T_{\rm cb}$ is still above $29\%$ for the electron densities from $0.8$ to $1.2$.

\begin{figure}[tb]
\resizebox{0.65\columnwidth}{!}{\rotatebox{0}{\includegraphics{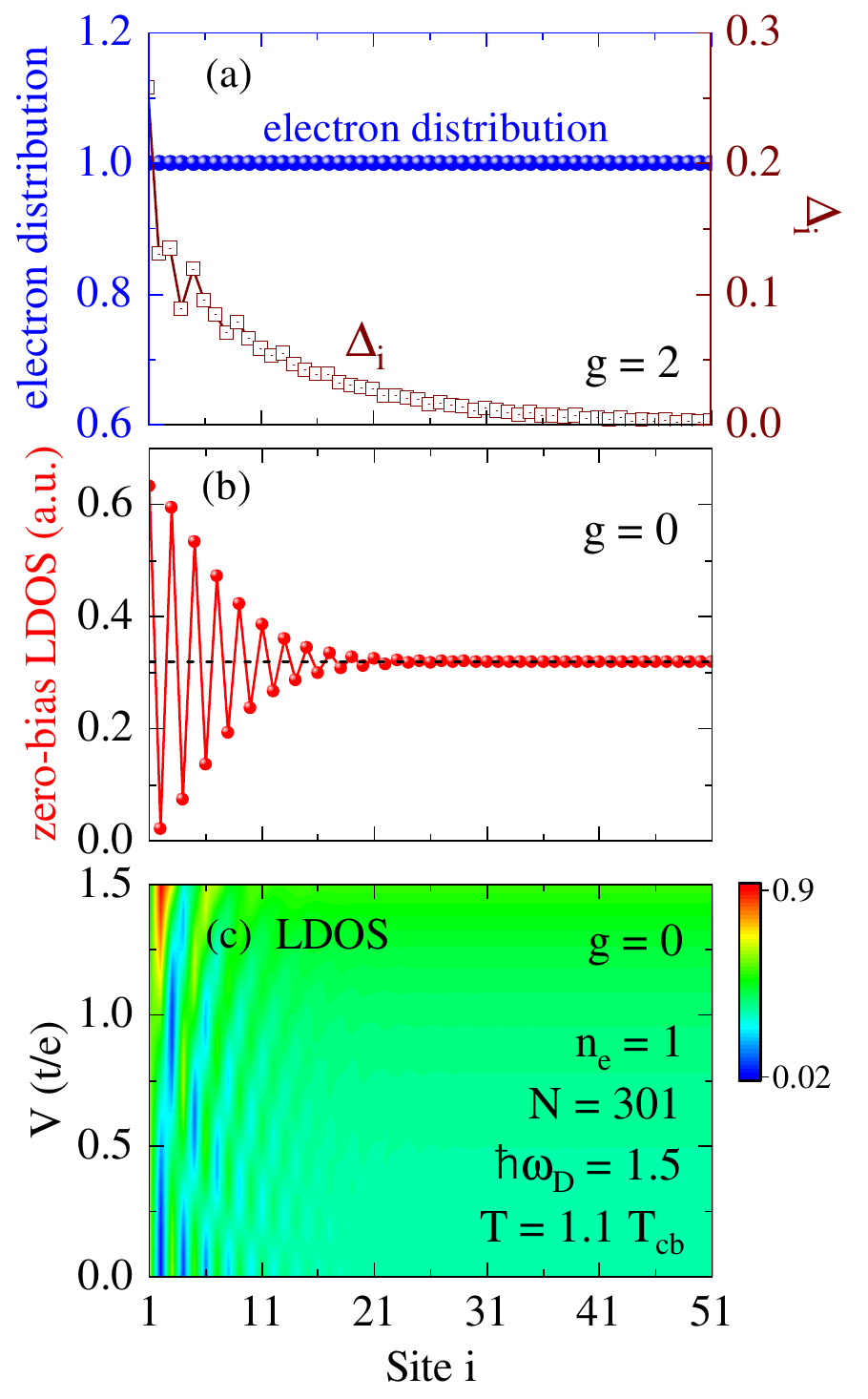}}}
\caption{\textcolor{black}{(a) The site-dependent order parameter and local electron density at $T=1.1T_{\rm cb}, n_e=1,\hbar\omega_D=1.5$, and $g=2$; (b) and (c) demonstrate the zero-bias and energy dependent normal LDOS ($g=0$) for the same $T$ and $n_e$ as in panel (a).}} \label{fig6}
\end{figure}

\textcolor{black}{It is important to stress that the obtained results for the surface superconductivity cannot be explained by an increase of the local electron density and the normal local DOS (LDOS) near the sample boundaries. To go in a more detail on this point, Fig.~\ref{fig6} demonstrates the site-dependent electron density and normal LDOS together with the order parameter near the left edge of the chain. From Fig.~\ref{fig6}(a) one can see that the surface effect in question is not related to a rise of the local electron density near the sample edges since the density is uniform. Indeed, the Friedel oscillations, present in the local density near the chain edges beyond the half-filling regime, are weakened and washed out when $n_e$ approaches $1$, as is seen from Fig.~\ref{fig6}(a). We find that the local density of electrons is constant when the surface effect is most pronounced.} 

\textcolor{black}{The Friedel oscillations in the normal ($g=0$) LDOS near the sample edges are present even in the half-filling case, as is seen from Fig.~\ref{fig6}(b) and (c). However, there is no any overall enhancement of the normal LDOS near the sample boundaries. Moreover, one can see that the Friedel oscillations are significant only in the domain with $i < 21$. They are completely washed out for $i > 25$. However, the surface-enhanced order parameter is not zero even for $i=51$ while the bulk order parameter is already zero in this case (we recall that here $T=1.1T_{\rm cb}$). Thus, one can conclude that the Friedel oscillations of the LDOS and electron density near the chain edges cannot explain the surface enhancement of the superconducting condensate. This confirms our conclusion that the surface enhancement of superconductivity found in the attractive Hubbard model for the Debye energies less than the band width is a result of the constructive interference of the pair states. This is similar to the results obtained previously~\cite{Croitoru2020} for the same model with $\hbar\omega_{\rm D} \gg 2$.}

Here the question may arise whether the mean field results obtained in the present study are reliable since 1D systems suffer from strong superconducting fluctuations~\cite{Efetov1974, Gorkov1975, Klemm1976, Schulz1983, Jerome1980}. The point is that the interference effects are not sensitive to the system dimensionality and the surface enhancement of superconductivity occurs also in 2D and 3D systems (see e.g. Ref.~\onlinecite{Samoilenka2020}), where the fluctuations are much less important. This is why we can expect that our conclusions obtained for the 1D chain (with a relatively simple formalism) are general and hold for higher dimensions. \textcolor{black}{For example, the surface superconductivity impact in a 3D superconductor occupying the half-space (say, for $x > 0$) can be estimated by introducing an additional factor in Eq.~(\ref{op}) that accounts for the states in the $y$ and $z$ directions (considering that these states are the plain waves). This changes the total DOS at the Fermi level but does not alter the constructive interference of the BdG pair states.}  

\textcolor{black}{In addition, since the interference of pair states can be influenced by the boundary conditions at the chain edges, it is necessary to say a few words about their possible effects in the context of the stability of the surface enhancement of superconductivity. The study performed in Ref.~\onlinecite{Croitoru2020} has demonstrated that the surface superconductivity is more sensitive to impurities than the bulk one. This is the reflection of the fact that the interference of the pair states is the origin of the surface enhancement. However, the effect survives at moderate surface disorder (roughness) unless the surface impurity potential becomes of the order of the hopping parameter. Further investigations of the boundary effects, including more sophisticated variants of the confinement potential at the sample boundaries, would be a significant deviation from the goals of the present study. Our consideration of the infinite potential walls at the chain edges~(open boundary conditions) is dictated by the fact that the recent results for the interference-induced surface superconductivity were obtained for infinite confinement barriers~\cite{Samoilenka2020, Croitoru2020}. Thus, our choice makes it possible to avoid any effects of a more elaborated finite potential when comparing our results with the earlier calculations.} 

As it follows from the present investigation, controlling the Debye frequency can be important to increase the superconducting surface temperature effect for the phonon-mediated superconductors. The Debye frequency depends on the phonon group velocity. There are several ways of controlling/tuning the phonon dispersion relation (phonon engineering~\cite{Balandin2012}) and hence the phonon group velocity. For example, by properly selecting the parameters of cladding materials and their thicknesses, one can control the group velocity of phonons near the sample surface~\cite{Pokatilov2005, Balandin2007}. In addition, the frequency and group velocity of acoustic phonons can decrease nonmonotonically with an increasing doping concentration, revealing pronounced phonon softening effects governed by the doping level~\cite{Guzman2022}. The phonon hardening can be reached by isotope substitutions like in H$_3$S, where replacement of $^{32}$S atoms by the heavier isotopes $^{33}$S, $^{34}$S, $^{35}$S, and $^{36}$S produces a significant effect on the lattice dynamics~\cite{Zheleznyak2018}. Finally, the strain at the sample surface/interface also affects the phonon structure and dispersion relation~\cite{Dickey1968,Naugle1973, Zhang2005, Lanzillo2014} and so, it can be used to manipulate the Debye frequency. Notice that the phonon softening near surfaces can have a dual effect on the surface superconductivity enhancement: firstly, by increasing the electron-phonon coupling, and secondly, by increasing $T_{\rm cs}$ as compared to $T_{\rm cb}$ due to changing the ratio of the Debye energy to the energy band width. Thus, taking into account the present technological possibilities of manipulating the Debye frequency in a controllable way, our research suggests an innovative way of tailoring the surface superconducting characteristics.

\begin{acknowledgments}
This work was supported by Natural Science Foundation of Zhejiang Province (Grants No. LY18A040002), Science Foundation of Zhejiang Sci-Tech University(ZSTU) (Grant No. 19062463-Y), Open Foundation of Key Laboratory of Optical Field Manipulation of Zhejiang Province(ZJOFM-2020-007). The study has also been funded within the framework of the HSE University Basic Research Program.
\end{acknowledgments}

\bibliographystyle{apsrev4-1}

\end{document}